\documentclass[preprint,aps,prd,nopacs,nofootinbib,11pt]{revtex4}

\usepackage{epsf,epsfig,psfrag,graphics,color}
\usepackage{amsmath}
\usepackage{color}
\usepackage{mathrsfs}
\usepackage{graphicx}
\usepackage{latexsym}
\usepackage{ulem}
\usepackage{graphicx}
\usepackage{subfigure}

\def\red{\textcolor[rgb]{1.00,0,0}}
\def\vec{\mathbf}

\begin{document}
\title{Relativistic corrections to light-cone distribution amplitudes of S-wave $B_c$ mesons and heavy quarkonia}
\author{Wei Wang$^1$~\footnote{E-mail: wei.wang@sjtu.edu.cn}, Ji Xu$^1$~\footnote{E-mail: xuji1991@sjtu.edu.cn}, Deshan Yang$^2$~\footnote{E-mail: yangds@ucas.ac.cn}, and Shuai Zhao$^1$~\footnote{E-mail: shuai.zhao@sjtu.edu.cn}}
\affiliation{$^1$INPAC, Shanghai Key Laboratory for Particle Physics and Cosmology,\\
School of Physics and Astronomy, Shanghai Jiao Tong University, Shanghai 200240, China\\
$^2$School of Physical Sciences, University of Chinese Academy of Sciences, Beijing 100049, China }
%\affiliation{Institute of High Energy Physics,
%Chinese Academy of Sciences, Beijing 100049, China}

%\date{March 29, 2017}

\vspace{1cm}
\begin{abstract}
In collinear factorization, light-cone distribution amplitudes
(LCDAs) are key ingredients to calculate the production rate   of a
hadron in high energy exclusive processes. For  a doubly-heavy meson
system (such as $B_c, J/\psi, \Upsilon$ etc), the LCDAs contain
perturbative scales
 that can be integrated out and then are
%{\color{blue}\sout{expressed in terms of} }
{\color{black} re-factorized into products of perturbatively calculable distribution parts and} non-relativistic QCD  matrix elements.
In this re-factorization scheme, the   LCDAs    are known at   next-to-leading order   in  the strong coupling constant
 $\alpha_s$ and  at  leading order  in  the velocity expansion. In this work,  we calculate the ${\cal O}( { v}^2)$ corrections to   twist-2 LCDAs  of S-wave $B_c$ mesons.
These  results are applicable to heavy quarkonia like $\eta_{c,b}$, $J/\psi$ and
$\Upsilon$ by setting $m_b=m_c$. We apply these relativistically
corrected LCDAs to study their inverse moments and a few Gegenbauer
moments which are important for phenomenological study. We point out that the relativistic corrections are
sizable, and comparable with the next-to-leading order radiative
corrections. These results for LCDAs are useful  in future
theoretical analyses of the productions %{\color{blue} \sout{and decays}}
 of heavy
quarkonia and $B_c$ mesons.
\end{abstract}
%\begin{description}
%\item[Keywords:]
%QCD Phenomenology, NLO Computations
%\end{description}

% 12.38.-t   Quantum chromodynamics
% 12.38.Cy   Summation of perturbation theory
% 12.39.St  Factorization
% 14.40.Gx   Mesons with S=C=B=0, mass > 2.5 GeV (including quarkonia)

\maketitle

%\newpage
%\tableofcontents

\section{Introduction\label{sect:Introduction}}

A typical  high energy process  with hadrons involves physics from  high scales, such as the center-of-mass collision energy, down to very low energy scales such as the   mass of a proton.  Without disentangling   physics associated with these scales, it is nearly hopeless to obtain reliable theoretical predictions for any such process.  The disentanglement is often referred to as factorization.

There exist two sorts of factorization schemes to deal with  hard
exclusive processes involving heavy quarkonia and $B_c$ mesons, if we regard heavy quarkonia and $B_c$ mesons as the non-relativistic bound-states of heavy quark and anti-quark. One is the non-relativistic QCD
(NRQCD) factorization \cite{Bodwin:1994jh}, in which the amplitude can be factorized into a product of the
perturbatively calculable short-distance coefficient and the
non-perturbative NRQCD matrix-element. For recent reviews of this
approach, one can refer to
Refs.~\cite{Brambilla:2004jw,Brambilla:2004wf,Brambilla:2010cs}. The
other one is called the collinear factorization, which has already been
established  for a few decades \cite{Lepage:1980fj,Chernyak:1983ej}.
In this framework, the amplitudes of hard exclusive processes can
be expressed as convolutions of the perturbatively calculable
hard-kernels and the universal light-cone distribution amplitudes
(LCDAs). It is necessary to mention at this moment that the
factorization approaches do not always hold, but instead should be
proved in a channel-by-channel manner. Once proved,  these
approaches can both clearly disentangle the short-distance and
long-distance contributions.

%{\color{blue} \sout{Processes involving}} {\color{black} The productions of}  heavy quarkonia and $B_c$ mesons %{\color{blue} \sout{at high energy}} {\color{black} in which the central collision energy is much higher than heavy-quark masses,} are    unique ones since
%{\color{blue} \sout{one may apply both}} {\color{black} the NRQCD and collinear} factorizations {\color{red} may be both applicable}. For instance,  the $B_c$ and $B_c^*$ mesons are  ground states of the heavy $\bar b c$  systems with the   Fock states having  the angular momentum

In literatures, the
collinear factorization and its extension $k_T$ factorization have
been adopted in
Refs.~\cite{Liu:2009qa,Liu:2010nz,Liu:2010da,Liu:2010kq,Xiao:2011zz,Rui:2015iia,Xiao:2013lia,Wang:2014yia,Liu:2017cwl,Wang:2016qli,Yang:2010ba}
to study various $B_c$ decays, where the LCDA of $B_c$ has
 been used to parametrize the non-perturbative physics in the system.
  Meanwhile the  NRQCD factorization has been adopted in Refs.~\cite{Qiao:2012vt,Qiao:2012hp,Shen:2014msa,Wang:2015bka,Zhu:2017lqu,Wu:2002ig,Chang:2005bf,Chang:2007si,Zheng:2017xgj},
  in which the non-perturbative physics reside in a few NRQCD matrix elements.
  The LHCb collaboration has started comprehensive experimental studies of the decays~\cite{Aaij:2014jxa},
  and some phenomenological studies can be found in Refs.~\cite{Wang:2008xt,Wang:2009mi,Sun:2008ew,Hwang:2008qi,Jia:2015pxx}.

Though the two factorization  approaches are different in nature, it has been argued that since the LCDAs of quarkonium and $B_c$ meson
 contain some perturbatively calculable scales like $\mathcal{O}(m^2)$, it is reasonable to consider the LCDAs can be further factorized
  into a product of a perturbatively calculable distribution part and a NRQCD matrix-element for the vacuum to quarkonium or $B_c$ state
  transition~\cite{Ma:2006hc} . This idea can be called as
  ``re-factorization". Some other approaches to bridge NRQCD and
  collinear factorization are proposed in
  Refs.~\cite{Jia:2008ep,Jia:2010fw}.

The re-factorization scheme has been extended to the next-to-leading
order ${\cal O}(\alpha_s)$ accuracy in
Refs.~\cite{Ma:2006hc,Bell:2008er,Wang:2013ywc,Xu:2016dgp}, and the
relation between the moments of LCDAs for heavy quarkonia and NRQCD
matrix elements are
known~\cite{Braguta:2006wr,Braguta:2007fh,Braguta:2007tq,Braguta:2008qe}.
However, the factorization is only discussed at leading order of the
velocity expansion $v$. In fact, the relativistic corrections are
also important for the calculation of LCDAs of quarkonium and $B_c$
meson. Relativistic corrections are often characterized by the relative
velocity of heavy quarks within the bound states. It is estimated as $v^2\sim 0.3$ for $\bar c c$ system which is sizable. Thus,
in order to test the applications of LCDAs in quarkonium and $B_c$
meson productions precisely,
  one should consider the relativistic corrections, which are the main focus of this paper. To do so,
   we will calculate the three relativistic corrected LCDAs for the S-wave $B_c$ mesons,
   and  the corresponding results for quarkonia like $J/\psi$ can be easily deduced by setting $m_b=m_c$.

The rest of this  paper is organized as follows: in section \ref{sect:Definitions}, we give the notations used in this paper, and present the definitions of the leading-twist LCDAs of the S-wave $B_c$ mesons in terms of the matrix-elements of a certain class of non-local QCD operators. In section \ref{sect:Calculations}, we present our results of the LCDAs at $\mathcal{O}(\alpha_s^0 v^0)$, $\mathcal{O}(\alpha_s^0 v^2)$ and $\mathcal{O}(\alpha_s^1 v^0)$;
 In section \ref{sect:Comparison}, we will apply the results for the
 LCDAs and study the inverse moment and Gegenbauer moments of LCDAs.
 A comparison of relativistic corrections and radiative corrections is made.
   Finally, we summarize our work in section \ref{sect:Summary}.

%%%%%%%%%%%%%%%%%%%%%%%%%%%%%%
\section{LCDAs for S-wave $B_c$ mesons\label{sect:Definitions}}
%%%%%%%%%%%%%%%%%%%%%%%%%%%%%%

%%%%%%%%%%%%%%%%%%%%%%%%%%%%%%
\subsection{Light-cone coordinates}
%%%%%%%%%%%%%%%%%%%%%%%%%%%%%%

As we will see below, LCDAs are defined in the light-cone coordinate frame, in which a four-vector $a$ can be expressed as
\begin{eqnarray}
a^\mu= n_+\cdot a \frac{n_-^\mu}{2}+n_-\cdot a
\frac{n_+^\mu}{2}+a_\perp^\mu.
\end{eqnarray}
Here $n_+$ and $n_-$ are two light-like vectors, with $n_\pm^2=0$,
$n_+\cdot n_-=2$ and $n_{\pm}\cdot a_\perp=0$.

%%%%%%%%%%%%%%%%%%%%%%%%%%
\subsection{Definition of  LCDAs}
%%%%%%%%%%%%%%%%%%%%%%%%%%
 To define  the  LCDAs,  one needs to introduce the gauge invariant non-local
quark bilinear operators:
\begin{eqnarray}
   J[\Gamma](\omega)\equiv (\bar b W_c)(\omega n_+){n}\!\!\!\slash_+\Gamma(W_c^\dag c)(0)\,,
\end{eqnarray}
where $b$ and $c$ are the quark fields. The Wilson-line
\begin{eqnarray}
W_c(x)=\textrm{P  exp} \Big( ig_s\int_{-\infty}^0 ds~ n_+A(x+sn_{+})
\Big) \nonumber
\end{eqnarray}
is a path-ordered exponential with the path along the $n_+$
direction and  $g_s$ is QCD coupling constant and $A_\mu\equiv
A_\mu^a(x)T^a$, $T^a$ are the generators of SU(3) group in the
fundamental representation.

The  leading-twist, i.e. twist-2,   LCDAs can be defined from matrix
elements of operator $J[\Gamma](\omega)$:
\begin{subequations}
\begin{eqnarray}
\langle B_c(^1S_0,P)\vert J[\gamma_5](\omega)\vert 0\rangle&=&- i f_{P}n_+P\int_0^1 dx ~e^{i\omega n_+Px} \hat{\phi}_{P}(x;\mu)\,, \\
\langle B_c(^3S_1,P,\epsilon^*)\vert J[1](\omega)\vert 0\rangle&=& f_{V}m_V n_+\epsilon^*\int_0^1dx~e^{i\omega n_+Px} \hat{\phi}_{V}^{\parallel}(x;\mu)\,,\\
\langle B_c(^3S_1,P,\epsilon^*)\vert
J[\gamma^\alpha_\perp](\omega)\vert 0\rangle&=&f_{V}^{\perp}n_+P
\epsilon_\perp^{*\alpha} \int_0^1 dx ~e^{i\omega n_+Px}
\hat{\phi}_{V}^{\perp}(x;\mu)\,.
\end{eqnarray}\label{eq:def1}
\end{subequations}
Here $f_P$, $f_V$ and $f^{\perp}_V$ are decay constants, $P$ is the momentum of the $B_c$ meson,
$\epsilon^*$ is the polarization vector for the vector $B_c$
meson. $\hat \phi_P(x)$, $\hat \phi_V(x)$ and $\hat
\phi_V^{\perp}(x)$ are LCDAs for pseudo-scalar, longitudinally polarized and transversely polarized vector $B_c$
mesons, respectively. $x$ denotes the light-cone momentum fraction and $\bar
x\equiv 1-x$. $\mu$ is the renormalization scale. Note that the state $\vert B_c\rangle$ in Eq.~(\ref{eq:def1}) is relativistically normalized.

The LCDAs are normalized as
\begin{eqnarray}
&&\int_0^1 dx \hat\phi_{P}(x)=\int_0^1 dx \hat\phi_{V}^\parallel (x)=\int_0^1 dx \hat\phi_{V}^\perp(x)=1\,.\label{condition1}
\end{eqnarray}

In the following analysis,  it is convenient to employ
the Fourier transformed form of Eq.~(\ref{eq:def1})
\begin{subequations}
\begin{eqnarray}
  \langle B_c(^1S_0,P)\vert Q[\gamma_5](x)\vert 0\rangle&=&- i f_{P} \hat{\phi}_{P}(x)\,,\\
\langle B_c(^3S_1,P,\epsilon^*)\vert Q[1](x)\vert 0\rangle&=& f_{V} \frac{m_V n_+\epsilon^*}{n_+P}\hat{\phi}_{V}^{\parallel}(x)\,,\\
\langle B_c(^3S_1,P,\epsilon^*)\vert Q[\gamma^\alpha_\perp](x)\vert
0\rangle&=&  f_{V}^{\perp} \epsilon_\perp^{*\alpha}
\hat{\phi}_{V}^{\perp}(x)\,,
\end{eqnarray}\label{eq:def2}
\end{subequations}
where the Fourier-transformed operator
\begin{eqnarray}
 Q[\Gamma](x)&\equiv&\big[      (\bar b W_c)(\omega n_+){n}\!\!\!\slash_+\Gamma(W_c^\dag c)(0)       \big]_{\rm F.T. }   =\int\frac{d\omega}{2\pi}e^{-ixn_+P\omega}(\bar b W_c)(\omega n_+){n}\!\!\!\slash_+\Gamma(W_c^\dag c)(0) .
\end{eqnarray}

It is worth noting that the operator $Q[\Gamma](x)$ is invariant under the longitudinal boost, i.e. $n_+\to \alpha n_+$ and $n_-\to \alpha^{-1} n_-$ where $\alpha$ is an arbitrary positive real number. This leads to the conclusion that both the decay constants and the LCDAs of $B_c$ mesons defined in Eq.~(\ref{eq:def2}) are also boost-invariant. This allows us to calculate these decay constants and LCDAs in any reference frame. In this paper, we will choose the rest frame of $B_c$ mesons, i.e
\begin{eqnarray}\label{eq:restframe}
P^\mu=(m_{B_c},0,0,0), ~~~ n_\pm^\mu=(1,0,0,\pm 1)\,,
\end{eqnarray}
which is particularly convenient for matching the decay constants and LCDAs to NRQCD matrix-elements.
%%%%%%%%%%%%%%%%%%%%
\subsection{Re-factorization of the LCDAs \label{sect:ref}}
%%%%%%%%%%%%%%%%%%%%

The re-factorization idea has been proposed  in Refs.~\cite{Ma:2006hc,Wang:2013ywc,Bell:2008er,Xu:2016dgp}, where all of the LCDAs of quarkonia or $B_c$ mesons can be factorized into products of perturbatively calculable distribution parts and non-perturbative NRQCD matrix elements. It means that, at the operator level, we have the generic matching equation
\begin{eqnarray}
  Q[\Gamma](x,\mu)=\sum\limits_{n=0}^\infty \frac{d^\Gamma_n (x,\mu)}{M^{n+1}} O_{\Gamma,n}^{\rm NRQCD} \,,
\end{eqnarray}
where $n$ denotes the order of $v$-expansion, $d^\Gamma_n (x,\mu)$ is the short-distance coefficient
 as a distribution over the  momentum fraction $x$, $ O_{\Gamma,n}^{\rm NRQCD}$ is the NRQCD operator which scales $\mathcal{O}(v^n)$ in the   power-counting, and scale $M\equiv m_b+m_c$ is introduced to balance the mass dimensions of the NRQCD operators so that the short-distance coefficients $d^{\Gamma}_n(x,\mu)$ are set to be dimensionless. Therefore, the LCDAs of $B_c$ meson can be expressed as
\begin{eqnarray}\label{eq:matching}
  \langle B_c\vert Q[\Gamma](x,\mu)\vert 0\rangle\simeq \sum\limits_{n=0}^\infty \frac{d^\Gamma_n(x,\mu)}{M^{n+1}}\langle B_c\vert O_{\Gamma,n}^{\rm NRQCD} \vert 0\rangle \,.
\end{eqnarray}

 In the present work, since we focus on the LCDAs of the S-wave $B_c$ mesons, up to ${\cal O}(v^2)$, the NRQCD operators that we will consider are
\begin{subequations}
\begin{eqnarray}
{O}_0(^1S_0)&\equiv& \psi_{b}^\dagger
\chi_{c}\,,\\[0.2cm]
{O}_0(^3S_1)&\equiv&\psi_{b}^\dagger
 \boldsymbol{\sigma}\cdot\boldsymbol{\epsilon} \chi_{c}\,,\\
{O}_2(^1S_0)&\equiv&  \psi_b^\dagger \left(-\frac{i}{2}\stackrel{\leftrightarrow}{\mathbf{D}}\right)^2 \chi_c \,,\\
{O}_2(^3S_1)&\equiv& \psi_b^\dagger\left
(-\frac{i}{2}\stackrel{\leftrightarrow}{\mathbf{D}}\right)^2
\boldsymbol{\sigma}\cdot\boldsymbol{\epsilon}\chi_c \,.
\end{eqnarray}\label{eq:3}
\end{subequations}
Here $\psi_b$ and $\chi_c$ are the two-component effective fields for the $b$ quark and $\bar c$ quark in the NRQCD, respectively, and $\psi_{b}^\dagger
\stackrel{\leftrightarrow}{\mathbf{D}}\chi_{c}\equiv
\psi_{b}^\dagger( \mathbf{D}\chi_{c} )-( \mathbf{D} \psi_{b}
)^\dagger\chi_{c}$ with $ \mathbf{D} ={\boldsymbol
\nabla} -ig_s  \mathbf{A}$.

Thus, up to $\mathcal{O}(v^2)$, we have the matching equations
\begin{subequations}
\begin{eqnarray}
%\langle B_c(^1S_0,P)\vert Q[\gamma_5](x)\vert 0\rangle{\nonumber}
- i f_{P} \hat{\phi}_{P}(x)&=&
 \frac{d^P_0(x)}{M}\langle
B_c(^1S_0,P)| {O}_0(^1S_0)|0\rangle+\frac{d^P_2(x)}{M^3}\langle
  B_c(^1S_0,P)|{O}_2(^1S_0)|0\rangle ,\\
%\langle B_c(^3S_1,P,\epsilon^*)\vert Q[1](x)\vert 0\rangle
f_{V} \hat{\phi}_{V}^{\parallel}(x)
&=& %\bigg(
\frac{d^V_0(x)}{M}\langle
B_c(^3S_1,P,\boldsymbol{\epsilon}^*)|{O}_0(^3S_1)|0\rangle+\frac{d^V_2(x)}{M^3}\langle
  B_c(^3S_1,P,\boldsymbol{\epsilon}^*)|{O}_2(^3S_1) |0\rangle%\bigg)n_+\epsilon^*
  ,\\
%\langle B_c(^3S_1,P,\epsilon^*)\vert Q[\gamma_\perp^\alpha](x)\vert 0\rangle
f_{V}^{\perp} \hat{\phi}_{V}^{\perp}(x)&=&\frac{
 d^{V_\perp}_0(x)}{M}\langle B_c(^3S_1,P,\boldsymbol{\epsilon}^*)|{O}_0(^3S_1)|0\rangle+\frac{d^{V_\perp}_2(x)}{M^3}\langle
  B_c(^3S_1,P,\boldsymbol{\epsilon}^*)|{O}_2(^3S_1) |0\rangle\,,
\end{eqnarray}\label{non-local-exp}
\end{subequations}
Here the  $\boldsymbol{\epsilon}^*$ is the polarization vector of
$^3S_1$ state, and $\boldsymbol{\epsilon}\cdot
\boldsymbol{\epsilon}^*=1$. $d^{P,V,V_\perp}_{i}(x)~ (i=0,2)$ are perturbatively calculable
Wilson coefficients. We should note that the matrix-elements of the
NRQCD operators in Eq.~(\ref{non-local-exp}) are
relativistically normalized.

By integrating over $x$ in Eq.~(\ref{non-local-exp}), and by imposing the normalization conditions given in Eq.~(\ref{condition1}), we can get the matching equations for the decay constants
\begin{subequations}
\begin{eqnarray}
%\langle B_c(^1S_0,P)\vert Q[\gamma_5](x)\vert 0\rangle{\nonumber}
- i f_{P} &=&
 \frac{C^P_0}{M}\langle
B_c(^1S_0,P)| {O}_0(^1S_0)|0\rangle+\frac{C^P_2}{M^3}\langle
  B_c(^1S_0,P)|{O}_2(^1S_0)|0\rangle ,\\
%\langle B_c(^3S_1,P,\epsilon^*)\vert Q[1](x)\vert 0\rangle
f_{V}
&=& %\bigg(
\frac{C^V_0}{M}\langle
B_c(^3S_1,P,\epsilon^*)|{O}_0(^3S_1)|0\rangle+\frac{C^V_2}{M^3}\langle
  B_c(^3S_1,P,\epsilon^*)|{O}_2(^3S_1) |0\rangle%\bigg)n_+\epsilon^*
  ,\\
%\langle B_c(^3S_1,P,\epsilon^*)\vert Q[\gamma_\perp^\alpha](x)\vert 0\rangle
f_{V}^{\perp} &=&
 \frac{C^{V_\perp}_0}{M}\langle B_c(^3S_1,P,\epsilon^*)|{O}_0(^3S_1)|0\rangle+\frac{C^{V_\perp}_2}{M^3}\langle
  B_c(^3S_1,P,\epsilon^*)|{O}_2(^3S_1) |0\rangle\,,
\end{eqnarray}\label{decay constant get}
\end{subequations}
with the short-distance coefficients
\begin{eqnarray}
	C_i^{\Gamma}&=&\int_0^1 dx \,d_i^{\Gamma}(x)\,,~~~~i=0,2\,;~~~\Gamma=P,V,V_\perp\,.
\end{eqnarray}
Then, the LCDAs $\hat{\phi}(x)$ can be derived straight-forwardly.
In the following, it is more convenient to express the LCDAs in the following expansions
\begin{subequations}
\begin{eqnarray}
 \hat \phi_P(x)&=&   \hat \phi_P^{(0,0)}(x)+  \hat \phi_P^{(1,0)}(x)+  \hat \phi_P^{(0,1)}(x), \\
 \hat \phi_V^{||}(x)&=&   \hat \phi_V^{||(0,0)}(x)+  \hat \phi_V^{||(1,0)}(x)+ \hat \phi_V^{||(0,1)}(x), \\
 \hat \phi_V^{\perp}(x)&=&   \hat \phi_V^{\perp(0,0)}(x)+  \hat \phi_V^{\perp(1,0)}(x)+ \hat \phi_V^{\perp(0,1)}(x)\,,
\end{eqnarray}
\end{subequations}
where the superscript $(i,j)$ %{\color{blue}\sout{where $i,j = 0, 1$}}
denotes the order of $\alpha_s$ and $v^2$-expansion. $\hat{\phi}^{(0,0)}(x)$ and $\hat{\phi}^{(1,0)}(x)$ have been given in Ref.~\cite{Xu:2016dgp}. The explicit expressions of $\hat{\phi}^{(0,1)}(x)$ which are the main results of this work will be presented in the next section.

%%%%%%%%%%%%%%%%%%%%%%%%%%%%%%
\section{Calculations of LCDAs for the S-wave $B_c$ mesons\label{sect:Calculations}}
%%%%%%%%%%%%%%%%%%%%%%%%%%%%%%
\subsection{Perturbative Matching}

As we describe in the previous section, to get the decay constants and LCDAs of S-wave $B_c$ mesons up to ${\cal O}(v^2)$, we need to obtain the short-distance coefficients $d_i^\Gamma(x)$ in Eq.~(\ref{non-local-exp}) and $C_i^\Gamma$ in Eq.~(\ref{decay constant get}) from  the perturbative matching.

%In this subsection, we will briefly describe the perturbative matching procedure to get the short-distance coefficients $d_i^\Gamma(x)$ and $C_i^\Gamma$ defined in Sect.~\ref{sect:ref}.

For the perturbative matching, one is allowed to choose any convenient process.
After calculating the matrix elements in both full theory and effective field theory, one can derive the short-distance coefficients by solving the matching equations. To do so, in the standard NRQCD matching procedure, we usually use the free $b$-quark and $\bar c$-quark pair to replace the corresponding $B_c$ meson state. Firstly, we calculate the corresponding on-shell amplitude in full QCD, then we expand the amplitude in terms of the relative momentum $q^i$ so that each expanded term has a definite scaling in $v$-expansion. Finally, we can extract the short-distance coefficients by identifying the corresponding NRQCD matrix-elements in the expanded amplitude.

We set the momenta for on-shell $b$-quark and $\bar c$-quark as
\begin{eqnarray}\label{eq:kin}
\left\{\begin{array}{l}
p_b^\mu=(E_b,{\bf q})\,,~~p_c^\mu=(E_c,-{\bf q})\,,~~P^\mu=p_b^\mu+p_c^\mu=(E,{\bf 0})\\
E_b=m_b+\frac{{\bf q}^2}{2 m_b}+{\cal O}(v^4)\,,~~~E_c=m_c+\frac{{\bf q}^2}{2 m_c}+{\cal O}(v^4)\,,\\
E=E_b+E_c=M+\frac{{\bf q}^2}{2 m_b}+\frac{{\bf q}^2}{2 m_c}+{\cal O}(v^4)\,.
\end{array}\right.
\end{eqnarray}
where $M\equiv m_b+m_c$, and we count $\vert {\bf q}\vert\sim {\cal O}(v)$.

At tree level,  we have the matrix element
\begin{eqnarray}%\label{eq:17}
&&\langle  b^a(p_b)\bar c^b(p_c)| Q[\Gamma](x)| 0  \rangle \nonumber\\
&=& \delta^{ab}\int\frac{d \omega}{2\pi}e^{-i(x-n_+p_b/n_+P)\omega n_+ P}\bar u_b(p_b){n}\!\!\!\slash_+\Gamma v_c(p_c)\nonumber\\
&=&\frac{\delta^{ab}}{n_+P}\delta\left(x-\frac{n_+p_b}{n_+P}\right)\bar u_b(p_b) {n}\!\!\!\slash_+\Gamma v_c(p_c)\,,\label{eq:treeME}
\end{eqnarray}
where the supscripts $a$ and $b$ are color indices for the
$b$ quark and $\bar c$ quark, respectively, and $n_\pm^\mu$ is set to have the explicit form in Eq.~(\ref{eq:restframe}).

Here it is worth addressing that the quark-states in Eq.~(\ref{eq:treeME}) are non-relativistically normalized. Thus, in the Dirac representation for Gamma matrices, the non-relativistically normalized 4-component on-shell spinors for $b$ quark and $\bar c$ quark can be written explicitly as
\begin{subequations}\label{2-component}
\begin{eqnarray}
u_{b}(p_b) &=&  \sqrt{\frac{E_b+m_b}{2E_b}}\left(
                                           \begin{array}{c}
                             \xi({\bf q}) \\
\frac{{\bf q}\cdot {\bf \sigma}}{E_b+m_b}\xi({\bf q})
                                           \end{array}
                                         \right)\,,\\
                                         v_{c}(p_c) &=&  \sqrt{\frac{E_c+m_c}{2E_c}}\left(
                                           \begin{array}{c}
-\frac{{\bf q}\cdot {\bf \sigma}}{E_b+m_c}\eta({-\bf q})\\
                             \eta({-\bf q})\end{array}
                                         \right)\,.
\end{eqnarray}
\end{subequations}
Here we suppress the helicity indices for spinors. $\xi({\bf q})$ and $\eta({\bf q})$ are the 2-component Pauli spinors for the $b$ quark and $\bar c$ quark, respectively, and they are related to the following NRQCD matrix-elements:
\begin{subequations}
	\begin{eqnarray}
	\langle 0\vert \psi_b^{a}(0)\vert b^b(p_b)\rangle &=& \delta^{ab}\xi({\bf q})\,,\\
	\langle \bar c^b(p_c) \vert \chi_c^{a}(0)\vert 0\rangle &=& \delta^{ab}\eta(-{\bf q})\,.
\end{eqnarray}
\end{subequations}
where the supscripts $a,b$ are color indices.

% Therefore, any 4-component spinor bilinear can be decomposed into a linear combination of four independent Pauli spinor bilinears
%\begin{eqnarray}
%	\bar u_b(p_b)\Gamma v_c(p_c)&=& a_0(\Gamma) \xi^\dag({\bf q})\eta(-{\bf q})+\sum\limits_{i=1}^3 a_3^i(\Gamma) \xi^\dag({\bf q}){\bf \sigma}^i\eta(-{\bf q})\,,
%\end{eqnarray}
%where the Pauli spinor billinear $\xi^\dag \eta$ and $\xi^\dag  \sigma^i\eta$ refer to "spin-singlet" and "spin-triplet", respectively.

For illustration, we take $\Gamma=\gamma_5$ in Eq.~(\ref{eq:treeME}) as an example to show how to match $Q[\gamma_5](x)$ to the NRQCD operators up to ${\cal O}(v^2)$. Explicitly, we have
\begin{eqnarray}
	&&\langle  b^a(p_b)\bar c^b(p_c)| Q[\gamma_5](x)| 0  \rangle=\frac{\delta^{ab}}{n_+P}\delta\left(x-\frac{n_+p_b}{n_+P}\right)\bar u_b(p_b) {n}\!\!\!\slash_+\gamma_5 v_c(p_c)\nonumber\\
	&=&\frac{\delta^{ab}}{E_b+E_c}\delta\left(x-\frac{E_b-q^3}{E_b+E_c}\right)\sqrt{\frac{(E_b+m_b)(E_c+m_c)}{2 E_b 2 E_c}}\nonumber\\
	&&\times\left\{\left[1-\left(\frac{q^3}{E_b+m_b}-\frac{q^3}{E_c+m_c}\right)-\frac{{\bf q}^2}{(E_b+m_b)(E_c+m_c)}\right]\xi^\dag({\bf q})\eta(-{\bf q})\right.\nonumber\\
	&&~~~~~~~~~\left.-i\epsilon_{3ij}q^i\left(\frac{1}{E_b+m_b}+\frac{1}{E_c+m_c}\right)\xi^\dag({\bf q})\sigma^j\eta(-{\bf q})\right\}\,.
	\end{eqnarray}
	Then we expand the above matrix-element up to the second power of $q^i$,
	\begin{eqnarray}\label{eq:qg5expansion}
	&&\langle  b^a(p_b)\bar c^b(p_c)| Q[\gamma_5](x)| 0  \rangle\nonumber\\
	&=&\frac{\delta^{ab}}{M}\xi^\dag({\bf q})\eta(-{\bf q})\left\{\delta\left(x-x_0\right)\left[1-\frac{q^3}{M}\frac{(1-2 x_0)}{2 x_0\bar x_0}-\frac{{\bf q}^2}{M^2}\frac{1 + 4 x_0 \bar x_0}{ 8x_0^2 \bar x_0^2}\right]\right.\nonumber\\
	&&\left.+\delta^\prime(x-x_0)\frac{q^3}{M}-\delta^\prime\left(x-x_0\right)\frac{(q^3)^2+{\bf q}^2}{M^2}\frac{1 - 2 x_0 }{ 2x_0 \bar x_0}+\delta^{\prime\prime}(x-x_0)\frac{(q^3)^2}{2M^2}\right\}\nonumber\\
	&&-i\frac{\delta^{ab}}{M}\frac{1}{2x_0\bar x_0}\xi^\dag({\bf q})\sigma^j\eta(-{\bf q})\epsilon_{3ij}\frac{q^i}{M}\left\{\delta\left(x-x_0\right)+\frac{q^3}{M}\delta^\prime(x-x_0)\right\}+{\cal O}(v^3)\,,
\end{eqnarray}
where $x_0\equiv m_b/(m_b+m_c)$ and $\bar x_0\equiv 1-x_0= m_c/(m_b+m_c)$.

In this work, we are only interested in the S-wave part of the above matrix-element. Due to the standard extraction procedure of the S-wave contribution in the literature, at first we neglect the first order expansion in $q^i$ (which is P-wave part that does not contribute to the vacuum to the S-wave state transition when only leading NRQCD interactions are considered), then replace $q^i q^j$ with ${\bf q}^2 \delta^{ij}/3$ in the second order expansion.

Thus, by identifying
\begin{subequations}
	\begin{eqnarray}
	\langle  b^a(p_b)\bar c^b(p_c)|
	O_0(^1S_0)| 0  \rangle &=&\delta^{ab}\xi^\dag({\bf q})\eta(-{\bf q})\,,\\
	\langle  b^a(p_b)\bar c^b(p_c)|
	O_2(^1S_0)| 0  \rangle &=&\delta^{ab}{\bf q}^2\xi^\dag({\bf q})\eta(-{\bf q})\,,
\end{eqnarray}
\end{subequations}
we get
\begin{eqnarray}
	&&\langle  b^a(p_b)\bar c^b(p_c)| Q[\gamma_5](x)| 0  \rangle^{\text{S-Wave}}\nonumber\\
	&=&\frac{1 }{M}\Bigg\{\delta\left(x-x_0\right)\langle  b^a(p_b)\bar c^b(p_c)| %\psi^\dag_b\chi_c
	O_0(^1S_0)| 0  \rangle\nonumber\\
	% &&-\left[\frac{(1-2 x_0)}{2 x_0\bar x_0}\delta\left(x-x_0\right)+\delta^\prime\left(x-x_0\right)\right]\frac{\langle  b^a(p_b)\bar c^b(p_c)| \psi^\dag_b\left(-\frac{i}{2}\overleftrightarrow{D}^3\right)\chi_c | 0  \rangle}{M}\nonumber\\
	%&&+\frac{1}{2x_0\bar x_0}\delta\left(x-x_0\right)\frac{\langle  b^a(p_b)\bar c^b(p_c)|i\epsilon_{3ij} \psi^\dag_b\left(-\frac{i}{2}\overleftrightarrow{D}^i\right)\sigma^j\chi_c | 0  \rangle}{M}\nonumber\\
	&&-\left[\frac{1 + 4 x_0 \bar x_0}{ 8x_0^2 \bar x_0^2}\delta(x-x_0)+\frac{2(1 - 2 x_0) }{3 x_0 \bar x_0}\delta^\prime\left(x-x_0\right)-\frac{\delta^{\prime\prime}(x-x_0)}{6}\right]\frac{\langle  b^a(p_b)\bar c^b(p_c)|
	%\psi^\dag_b\left(-\frac{i}{2}\overleftrightarrow{\bf D}\right)^2\chi_c
	O_2(^1S_0)| 0  \rangle}{M^2}\nonumber\\
	%&&+\left[\delta^\prime\left(x-x_0\right)\frac{1 - 2 x_0 }{ x_0 \bar x_0}+\delta^{\prime\prime}(x-x_0)\right]\frac{\langle  b^a(p_b)\bar c^b(p_c)| \psi^\dag_b\left[\left(-\frac{i}{2}\overleftrightarrow{D}^3\right)\left(-\frac{i}{2}\overleftrightarrow{D}^3\right)-\frac{1}{3}\left(-\frac{i}{2}\overleftrightarrow{\bf D}\right)^2\right]\chi_c | 0  \rangle}{2M^2}\nonumber\\
	&&%-\frac{1}{2x_0\bar x_0}\delta^\prime(x-x_0)\frac{i\epsilon_{3ij}\langle  b^a(p_b)\bar c^b(p_c)| \psi^\dag_b\left(-\frac{i}{2}\overleftrightarrow{D}^i\right)\left(-\frac{i}{2}\overleftrightarrow{D}^3\right)\sigma^j\chi_c | 0  \rangle}{M^2}
	+{\cal O}(v^3)\Bigg\}\,.
\end{eqnarray}
Therefore, we can easily extract the short-distance coefficients for S-wave operators defined in Eq.(\ref{non-local-exp}),
\begin{subequations}
	\begin{eqnarray}
	d_0^P(x)&=&\delta(x-x_0)\,,~~~\\
	d_2^P(x)&=&-\left[\frac{1 + 4 x_0 \bar x_0}{ 8x_0^2 \bar x_0^2}\delta(x-x_0)+\frac{2(1 - 2 x_0) }{ 3x_0 \bar x_0}\delta^\prime\left(x-x_0\right)-\frac{1}{6}\delta^{\prime\prime}(x-x_0)\right]\,,
\end{eqnarray}
\end{subequations}
and in turn,
	\begin{eqnarray}
	C_0^P=1\,,~~~
	C_2^P=-\left[\frac{1 + 4 x_0 \bar x_0}{ 8x_0^2 \bar x_0^2}\right]\,.
\end{eqnarray}

It is worth noting that the above matching procedure by directly decomposing the 4-component Dirac spinors to Pauli spinors is not very convenient and efficient in the cases that the Dirac structures become
complicated. In the literatures, the covariant projection
approach are commonly used~\cite{Kuhn:1979bb,Guberina:1980dc,Bodwin:2002hg,Wang:2015bka}.
Generally, one applies the replacements
\begin{subequations}
	\begin{eqnarray}
 v(p_c)\bar u(p_b) &\to &\Pi_0(p_b,p_c)=\frac{i}{2\sqrt{2 E_b
  E_c}\sqrt{(E_c+m_c)(E_b+m_b)}}\left(
  p\!\!\!\slash_c-m_c\right)\gamma_5\frac{
  P\!\!\!\!\slash+E_b+E_c}{2(E_b+E_c)}\left(p\!\!\!\slash_b+m_b\right)\,,\nonumber\\\\
    v(p_c)\bar u(p_b)  &\to & \Pi_1(p_b,p_c)=-\frac{1}{2\sqrt{2 E_b
  E_c}\sqrt{(E_c+m_c)(E_b+m_b)}}\left(
  p\!\!\!\slash_c-m_c\right) \epsilon\!\!\!\slash^*\frac{
  P\!\!\!\!\slash+E_b+E_c}{2(E_b+E_c)}\left( p\!\!\!\slash_b+m_b\right)\,,\nonumber\\\label{proj2}
\end{eqnarray}
\end{subequations}
to project out the spin-singlet and spin-triplet parts, respectively. After expanding the resulting amplitudes in the relative momentum ${\bf q}$, one can extract the S-wave contributions by neglecting the first order terms in ${\bf q}$ and making the replacement $q^i q^j\to {\bf q}^2 \delta^{ij}/3$. Finally one can obtain the S-wave short-distance coefficients by identifying the NRQCD matrix elements at tree level
\begin{subequations}\label{NRQCD:matrix:elements:LO}
\begin{align}
& \langle b \bar c({}^1S_0)|\psi_b^\dagger \chi_c | 0\rangle^{(0)} = \sqrt{2N_c},\\
 & \langle b \bar c({}^3S_1,{\bf \epsilon}^*)|\psi_b^\dagger\boldsymbol{\sigma\cdot \epsilon} \chi_c |
0\rangle^{(0)} = \sqrt{2N_c},\\
& \langle b \bar c({}^1S_0)
|\psi_b^\dagger(-\frac{i}{2}{\overleftrightarrow{ {\mathbf
D}}})^2\chi_c |0 \rangle^{(0)} %=  \langle b \bar c({}^1S_0)|\psi_b^\dagger(-\frac{i}{2}{\overleftrightarrow{ {\boldsymbol\nabla}}})^2\chi_c |0 \rangle^{(0)}
=\sqrt{2 N_c}\,{\bf q}^2,\\
& \langle b \bar c({}^3S_1,{\bf \epsilon}^*)
|\psi_b^\dagger(-\frac{i}{2}{\overleftrightarrow{ {\mathbf
D}}})^2\boldsymbol{\sigma\cdot \epsilon}\chi_c |0 \rangle^{(0)}
%=\langle b \bar c({}^3S_1,{\bf \epsilon}^*) |\psi_b^\dagger(-\frac{i}{2}{\overleftrightarrow{ {\boldsymbol\nabla}}})^2\boldsymbol{\sigma\cdot \epsilon}\chi_c |0 \rangle^{(0)}
= \sqrt{2 N_c}\,{\bf q}^2,
\end{align}
\end{subequations}
where $\vert b\bar c(^{2S+1}S_J)\rangle$ is formally a non-relativistically normalized color-singlet quark-anti-quark pair state. The factor $\sqrt{2N_c}$ is due to the spin and color factors of the normalized $b\bar c({}^{2S+1}S_J)$ state.

Since the Dirac structures involved in this work is not very complicated, we use both methods of the direct spinor decomposition and covariant projectors to do the matching for cross-checks. We get the exactly same results as it should be.

%%%%%%%%%%%%%%%%%%%%%%%%%%%%%%
\subsection{Results of the S-wave LCDAs up to $\mathcal{O}(\alpha_s^0 v^2)$}
%%%%%%%%%%%%%%%%%%%%%%%%%%%%%%
Following the matching procedure described in the previous subsection, we get the rest of the necessary short-distance coefficients up to ${\cal O}(v^2)$ at tree level,
\begin{subequations}
	\begin{eqnarray}
	d_0^V(x)&=&d_0^{V_\perp}(x)=\delta(x-x_0)\,,~~~\\
	d_2^V(x)&=&-\left[\frac{3 + 4 x_0 \bar x_0}{ 24x_0^2 \bar x_0^2}\delta(x-x_0)+\frac{2(1 - 2 x_0) }{ 3x_0 \bar x_0}\delta^\prime\left(x-x_0\right)-\frac{1}{6}\delta^{\prime\prime}(x-x_0)\right]\,,\\
	d_2^{V_\perp}(x)&=&-\left[\frac{3 + 8 x_0 \bar x_0}{ 24x_0^2 \bar x_0^2}\delta(x-x_0)+\frac{2(1 - 2 x_0) }{ 3x_0 \bar x_0}\delta^\prime\left(x-x_0\right)-\frac{1}{6}\delta^{\prime\prime}(x-x_0)\right]\,,
\end{eqnarray}
\end{subequations}
and consequently,
\begin{subequations}
	\begin{eqnarray}
	C_0^V&=&C_0^{V_\perp}=1\,,~~~\\
	C_2^V&=&-\frac{3 + 4 x_0 \bar x_0}{ 24x_0^2 \bar x_0^2}\,,\\
	C_2^{V_\perp}&=&-\frac{3 + 8 x_0 \bar x_0}{ 24x_0^2 \bar x_0^2}\,.
\end{eqnarray}
\end{subequations}

In turn, with Eqs.~(\ref{non-local-exp}) and (\ref{decay constant get}), at tree-level and up to ${\cal O}(v^2)$,
we have the decay constants in terms of NRQCD matrix elements as
\begin{subequations}
	\begin{eqnarray}
%\langle B_c(^1S_0,P)\vert Q[\gamma_5](x)\vert 0\rangle{\nonumber}
- i f_{P} &=&
 \frac{\langle
B_c(^1S_0,P)| {O}_0(^1S_0)|0\rangle}{M}\left\{1-\left[\frac{1 + 4 x_0 \bar x_0}{ 8x_0^2 \bar x_0^2}\right]\frac{ \langle \textbf{q}^2\rangle_P}{M^2}\right\}\,,\\
  f_{V} &=&
 \frac{\langle
B_c(^3S_1,P,{\bf \epsilon^*})| {O}_0(^3S_1)|0\rangle}{M}\left\{1-\left[\frac{3 + 4 x_0 \bar x_0}{ 24x_0^2 \bar x_0^2}\right]\frac{ \langle \textbf{q}^2\rangle_V}{M^2}\right\}\,,\\
f_{V}^\perp &=&
 \frac{\langle
B_c(^3S_1,P,{\bf \epsilon^*})| {O}_0(^3S_1)|0\rangle}{M}\left\{1-\left[\frac{3 + 8 x_0 \bar x_0}{ 24x_0^2 \bar x_0^2}\right]\frac{ \langle \textbf{q}^2\rangle_V}{M^2}\right\}\,,  \label{final decay constant
get3}
  \end{eqnarray}\label{final decay constant
get}
  \end{subequations}
and the LCDAs as
  \begin{subequations}
  \begin{eqnarray}
  \hat\phi_P^{(0,0)}(x)&=&\hat\phi_V^{\parallel(0,0)}(x)=\hat\phi_{V}^{\perp(0,0)}(x)=\delta(x-x_0)\,,\\
  \hat\phi_P^{(0,1)}(x)&=&-\frac{ \langle \textbf{q}^2\rangle_P}{M^2}\Big[ \frac{2(1-2x_0)}{3x_0\bar x_0}\delta'(x-x_0)-\frac{1}{6}\delta''(x-x_0) \Big]\,,\\
  \hat\phi_V^{\parallel (0,1)}(x)&=&\hat\phi_V^{\perp(0,1)}(x)=-\frac{ \langle \textbf{q}^2\rangle_V}{M^2}\Big[ \frac{2(1-2x_0)}{3x_0\bar x_0}\delta'(x-x_0)-\frac{1}{6}\delta''(x-x_0) \Big]\,,
  \end{eqnarray}
\end{subequations}
where $\langle \textbf{q}^2\rangle_{P,V}$ are the mean values of
$\textbf{q}^2$ in scalar and vector $B_c$ mesons respectively:
\begin{subequations}
 \begin{eqnarray}
  \langle \textbf{q}^2\rangle_P&\equiv&  \frac{\langle
  B_c(^1S_0,P)|{\psi^{\dag}_b\big(-\frac{i}{2}\overleftrightarrow{\mathbf{D}}\big)^2\chi_c}|0\rangle}{\langle B_c(^1S_0,P)| {\psi^{\dag}_b\chi_c}|0\rangle}\,, \\
  \langle \textbf{q}^2\rangle_V&\equiv& \frac{\langle
  B_c(^3S_1,P,{\bf \epsilon}^*)|{\psi^{\dag}_b\big(-\frac{i}{2}\overleftrightarrow{\mathbf{D}}\big)^2 \boldsymbol{\sigma}\cdot \boldsymbol{\epsilon}\chi_c} |0\rangle}{\langle B_c(^3S_1,P,{\bf \epsilon}^*)|{\psi^{\dag}_b\boldsymbol{\sigma}\cdot \boldsymbol{\epsilon}\chi_c}|0\rangle}\,,
 \end{eqnarray}
 \end{subequations}
which characterize ${\cal O}(v^2)$ relativistic corrections to the LCDAs.

By setting $m_b=m_c=m$, i.e. $x_0=1/2$, we recover the $f,~\hat
\phi(x)$ for heavy quarkonia at tree-level as given in
Ref.~\cite{Wang:2013ywc}; and also Eq.(\ref{final decay constant
get3}) returns to the result for $f_{J/\psi}^T$ in
Ref.~\cite{Bodwin:2014bpa}. Note that the matrix elements in
(\ref{final decay constant get}) are relativistically normalized
which are different from the widely-used non-relativistic
normalization. To get the results in the latter
form, one can multiply the factor $\sqrt{2 m_{B_c}}$ in the right hand side of Eq.~(\ref{final decay
constant get}) since it is assumed that the states are non-relativistically normalized, as shown
by, e.g., Eq.~(3) in Ref.~\cite{Bodwin:2008vp}.

%%%%%%%%%%%%%%%%%%%%%%%%%%%%%%
\subsection{Results of the S-wave LCDAs at $\mathcal{O}(\alpha_s^1 v^0)$} %next-to-leading order of the strong coupling and leading order of the velocity expansion
%%%%%%%%%%%%%%%%%%%%%%%%%%%%%%

${\cal O}(\alpha_s)$ radiative corrections are usually counted as important as ${\cal O}(v^2)$ relativistic corrections. For completeness, we quote the results of ${\cal O}(\alpha_s)$ radiative corrections from Ref.~\cite{Xu:2016dgp}.

We have
\begin{subequations}
\begin{eqnarray}
  C_0^{P}&=&    1+\frac{\alpha_s}{4\pi}C_F\left[3(x_0-\bar x_0)\ln{\frac{x_0}{\bar x_0}-6+4\Delta}  \right] \,,\\
 C_0^{ V}&=&   1+ \frac{\alpha_s}{4\pi}C_F\left[3(x_0-\bar x_0)\ln{\frac{x_0}{\bar x_0}}-8 \right] \,,\\
 C_0^{V_\perp}&=&  1+ \frac{\alpha_s}{4\pi}C_F \left[-\ln{\frac{\mu^2}{M^2}}-(3-8x_0)\ln x_0-(3-8\bar x_0)\ln \bar x_0-8 \right],
\end{eqnarray}
\end{subequations}
for the decay constants, and
\begin{subequations}
\begin{eqnarray}
 \hat \phi_P^{(1,0)}(x;\mu)&=&\frac{\alpha_s}{4\pi}C_F \bigg\{\Phi_1(x,x_0)
+8\Delta\left[\frac{x}{x_0}\theta(x_0-x)+\left(  x\leftrightarrow \bar x, \;\; x_0\leftrightarrow \bar x_0  \right)\right]_{+}\bigg\} \,, \\
%%%%%%%%%%%%%%%%%%%%%%%%%%%%%%%%%%%%%%%%%%%%%%%%%
 %%%%%%%%%%%%%%%%%%%%%%%%%%%%%%%%%%%%%%%%%%%%%
   \hat \phi_V^{\parallel(1,0)}(x;\mu)&=&\frac{\alpha_s}{4\pi}C_F \bigg\{\Phi_1(x,x_0)
-4\left[\frac{x}{x_0}\theta(x_0-x)+\left(  x\leftrightarrow \bar x, x_0\leftrightarrow \bar x_0   \right)\right]_{+}\bigg\} \,, \label{eq:INNRQCD}\\
 %%%%%%%%%%%%%%%%%%%%%%%%%%%%%%%%%%%%%%%%%%%%%%%%%%%
 \hat \phi_V^{\perp(1,0)}(x;\mu)&=&\frac{\alpha_s}{4\pi}C_F \bigg\{\Phi_1(x,x_0) \nonumber\\
&
-&2\left[\bigg( \ln{\frac{\mu^2}{M^2(x_0-x)^2}}-1\bigg)\bigg( \frac{x}{x_0}\theta(x_0-x)+\left( x\leftrightarrow \bar x,  x_0\leftrightarrow \bar x_0   \right) \bigg) \right]_+\bigg\} \,,
\end{eqnarray}
\end{subequations}
with
\begin{eqnarray}
\Phi_1(x,x_0)&=&2\left[ \left( \ln{\frac{\mu^2}{M^2(x_0-x)^2}}-1 \right)\left( \frac{x_0+\bar x}{x_0-x}\frac{x}{x_0}\theta(x_0-x)+\left( x\leftrightarrow \bar x,  x_0\leftrightarrow \bar x_0   \right) \right) \right]_+ \nonumber\\
&&~~+\left[ \frac{4x\bar x}{(x_0-x)^2}\right]_{++}+\left[ 4x_0\bar x_0 \ln{\frac{x_0}{\bar x_0}}+2(2x_0-1) \right]\delta^\prime(x-x_0)\,,
\end{eqnarray}
for the LCDAs. Here
$\Delta=0$ is for   NDR scheme \cite{Chanowitz:1979zu},
and
$\Delta=1$ for the HV scheme \cite{tHooft:1972tcz}.  The $++$- and $+$-distributions are defined as
\begin{subequations}
\begin{eqnarray}
\int _0^1dx\Big[ f(x)  \Big]_{++}g(x)&=&\int _0^1dx f(x)(g(x)-g(x_0)-g'(x_0)(x-x_0)),  \\
\int _0^1dx\Big[ f(x)  \Big]_{+}g(x)&=&\int _0^1dx
f(x)(g(x)-g(x_0))\,,
\end{eqnarray}
\end{subequations}
where $g(x)$ is a smooth test function.

%%%%%%%%%%%%%%%%%%%%%%%%%%%%%%%
\section{Phenomenological  Results \label{sect:Comparison}}
%%%%%%%%%%%%%%%%%%%%%%%%%%%%%%%

\subsection{Inverse Moment}
 In the production or  decay processes of mesons where collinear
factorization can be applied, the inverse moment  of the LCDA is crucial because
the hard kernel are often functions of $1/x$ and $1/\bar x$. Thus
it is of phenomenological importance to apply the results in last
section to study the inverse moment. The inverse moment of the
LCDAs is defined by
\begin{eqnarray}
\left\langle \frac{1}{x}\right\rangle_{\Gamma}\equiv \int_0^1
dx\frac{\hat\phi_\Gamma(x;\mu)}{x}\,.
\end{eqnarray}
The inverse moment can be calculated from the LCDA as given in the
above section. The results are listed below,
\begin{subequations}
 \begin{eqnarray}
\left\langle \frac{1}{x}\right\rangle_P &=&\frac{1}{x_0}\left\{1+\frac{\alpha_s}{4\pi}C_F\left[\left(3+2\ln x_0\right)\ln\frac{\mu^2}{M^2}-6\bar x_0\ln\bar x_0-6 x_0\ln x_0 -2\ln x_0-8\Delta\frac{x_0}{\bar x_0}\ln x_0\right.\right.\nonumber\\
&&~~~~~~\left.\left.+4\,{\rm Li}_2(x_0)-2 \ln^2 x_0-\frac{2\pi^2}{3}+6-4\Delta\right]
%\right.\nonumber\\&&~~~~~~\left.
-\frac{(1-3x_0)}{3x_0^2\bar x_0 }\frac{\langle \textbf{q}^2\rangle_P }{M^2}\right\} \,,\\
%%%%%%%%%%%%%%%
\left\langle \frac{1}{x}\right\rangle_V&=&\frac{1}{x_0}\left\{1+\frac{\alpha_s}{4\pi}C_F\left[\left(3+2\ln x_0\right)\ln\frac{\mu^2}{M^2}-6\bar x_0\ln\bar x_0-6 x_0\ln x_0 -2\ln x_0+4\frac{x_0}{\bar x_0}\ln x_0 \right.\right.\nonumber\\
&&\left.\left.~~~~~~ +4\,{\rm Li}_2(x_0)-2 \ln^2 x_0-\frac{2\pi^2}{3}+8\right]
%\right.\nonumber\\&&~~~~~~\left.
-\frac{(1-3x_0)}{3x_0^2\bar x_0}\frac{\langle \textbf{q}^2\rangle_V}{M^2}\right\} \,,\\
%%%%%%%%%%%%%%%
\left\langle \frac{1}{x}\right\rangle_{V_\perp}&=&\frac{1}{x_0}\left\{1+\frac{\alpha_s}{4\pi}C_F\left[\left(4+\frac{2}{\bar x_0}\ln x_0\right)\ln\frac{\mu^2}{M^2} +4(x_0-2)\ln\bar x_0-2(1+2x_0)\ln x_0-2\frac{x_0}{\bar x_0} \ln x_0\right.\right.\nonumber\\
&&~~~~~~ \left.\left.+\frac{1}{\bar x_0}\left(4\,{\rm Li}_2(x_0)-2\ln^2 x_0-\frac{2\pi^2}{3}\right)+8\right]
%\right.\nonumber\\&&~~~~~~\left.
-\frac{(1-3x_0)}{3x_0^2\bar x_0 }\frac{\langle \textbf{q}^2\rangle_V}{M^2}\right\}\,.
%%%%%%%%%%%%%%%
\end{eqnarray}
\end{subequations}
Here $\Delta=0$ is for the NDR scheme, and $\Delta=1$ is for the HV
scheme.

%%%%%%%%%%%%%%%%%%%%%%%%%%%%%%%%%%%%%%%%%%%%%%

\subsection{Gegenbauer moments}

The Gegenbauer moments are also commonly used, which are  defined by
\begin{eqnarray}
(a_n)_{\Gamma}\equiv\frac{2(2n+3)}{3(2+n)(1+n)}\int_0^1
dx\hat\phi_\Gamma(x)C_n^{(3/2)}(2x-1)\,.
\end{eqnarray}
With the results of LCDA derived above, one can work out
the first two Gegenbauer moments,
and  the results are given below.
 For the  $a_1$, we have
 \begin{subequations}\label{mm:a1}
\begin{eqnarray}
(a_1)_P &=&\frac{5}{3}(2x_0-1)\Bigg\{ 1+\frac{\alpha_s}{4\pi}C_F\frac{4}{9(2x_0-1)}\Big[ 6(1-2x_0)\ln{\frac{\mu^2}{M^2}}+3\bar x_0(3-7\bar x_0)\ln{\bar x_0}\nonumber\\
&&~~~~~~~~~~~~~~~~-3x_0(3-7x_0)\ln{x_0}+ (1-2x_0)(19+6\Delta)\Big]
%~~~~~~~~-3(x_0-1)(7x_0-4)\ln{\bar x_0}+3x_0(7x_0-3)\ln{x_0}
  %\Big]\nonumber\\&&~~~~~~~~~~~~~~~~~~~~~~~~
  -\frac{4}{3x_0\bar x_0}\frac{ \langle
\textbf{q}^2\rangle_P}{M^2}  \Bigg\}\,,\\
(a_1)_V &=&\frac{5}{3}(2x_0-1)\Bigg\{ 1+\frac{\alpha_s}{4\pi}C_F\frac{4}{9(2x_0-1)}\Big[  6(1-2x_0)\ln{\frac{\mu^2}{M^2}}+3\bar x_0(3-7\bar x_0)\ln{\bar x_0}\nonumber\\
%&&~~~~~~~~~~~~~~~~~~~~~~~~-3(x_0-1)(7x_0-4)\ln{\bar x_0}+3x_0(7x_0-3)\ln{x_0}  \Big]\nonumber\\
%&&~~~~~~~~~~~~~~~~~~~~~~~~-\frac{4}{3x_0\bar x_0M^2} \langle \textbf{q}^2\rangle_V  \Bigg\},
&&~~~~~~~~~~~~~~~~-3x_0(3-7x_0)\ln{x_0}+16(1-2x_0)\Big]
%~~~~~~~~-3(x_0-1)(7x_0-4)\ln{\bar x_0}+3x_0(7x_0-3)\ln{x_0}
  %\Big]\nonumber\\&&~~~~~~~~~~~~~~~~~~~~~~~~
  -\frac{4}{3x_0\bar x_0}\frac{ \langle
\textbf{q}^2\rangle_V}{M^2}  \Bigg\}\,,\\
(a_1)_{V_\perp} &=&\frac{5}{3}(2x_0-1)\Bigg\{ 1+\frac{\alpha_s}{4\pi}C_F\frac{1}{2x_0-1}\Big[  -2(1-2x_0)^2\ln{\frac{\mu^2}{M^2}}+4\bar x_0\ln{\bar x_0}\nonumber\\
&&~~~~~~~~~~~~~~~~+4x_0(4x_0-3)\ln{x_0} -8(1-2x_0)^2\Big]
%-4(x_0-1)\ln{\bar x_0}+4x_0(4x_0-3)\ln{x_0} \Big]\nonumber\\&&~~~~~~~~~~~~~~~~~~~~~~~~
-\frac{4}{3x_0\bar x_0} \frac{\langle \textbf{q}^2\rangle_V}{M^2}  \Bigg\}\,.
\end{eqnarray}
\end{subequations}
Notice that for heavy quarkonia, the first Gegenbauer moment $a_1$ vanishes, since the $a_1$ reflects the asymmetry in the momentum distribution.
For the $a_2$, we have
\begin{subequations}\label{mm:a2}
\begin{eqnarray}
(a_2)_P &=&\frac{7}{18}C_{2}^{(3/2)}(2x_0-1)\Bigg\{ 1+\frac{\alpha_s}{4\pi}C_F\frac{5}{36(1-5x_0\bar x_0)}\Big[ -30(1-5x_0\bar x_0)\ln{\frac{\mu^2}{M^2}}\nonumber\\
&&+12\bar x_0(5+x_0(-26+27x_0))\ln{\bar x_0}+12x_0(6+x_0(-28+27x_0))\ln{x_0}\nonumber\\
%-x_0(x_0-1)(589+120\Delta)-30(1+5x_0(x_0-1))\ln{\frac{\mu^2}{M^2}}{\nonumber}\\
&&+x_0\bar x_0(589+120\Delta) -119-24\Delta
%-12(x_0-1)(5+x_0(-26+27x_0))\ln{\bar x_0}+12x_0(6+x_0(-28+27x_0))\ln{x_0}
  \Big]%\nonumber\\&&-\frac{5(2+9x_0(x_0-1))}{3x_0\bar x_0(1+5x_0(x_0-1))}
  -\frac{5(2-9x_0\bar x_0)}{3x_0\bar x_0(1-5x_0\bar x_0)}\frac{ \langle \textbf{q}^2\rangle_P}{M^2}
  \Bigg\}\,,
\\
(a_2)_V &=&\frac{7}{18}C_{2}^{(3/2)}(2x_0-1)\Bigg\{ 1+\frac{\alpha_s}{4\pi}C_F\frac{5}{36(1-5x_0\bar x_0)}\Big[-30(1-5x_0\bar x_0)\ln{\frac{\mu^2}{M^2}} \nonumber\\
&&{\nonumber}\\
&&+12\bar x_0(5+x_0(-26+27x_0))\ln{\bar x_0}+12x_0(6+x_0(-28+27x_0))\ln{x_0}
%-12(x_0-1)(5+x_0(-26+27x_0))\ln{\bar x_0}+12x_0(6+x_0(-28+27x_0))\ln{x_0}
 \nonumber\\
  &&   -107+529x_0\bar x_0\Big]
  -\frac{5(2-9x_0\bar x_0)}{3x_0\bar x_0(1-5x_0\bar x_0)}\frac{ \langle \textbf{q}^2\rangle_V }{M^2}\Bigg\}\,,
\\
(a_2)_{V_\perp} &=&\frac{7}{18}C_{2}^{(3/2)}(2x_0-1)\Bigg\{ 1+\frac{\alpha_s}{4\pi}C_F\frac{10}{9(1-5x_0\bar x_0)}\Big[ -3(1-5x_0\bar x_0)\ln{\frac{\mu^2}{M^2}}\nonumber\\
&&+3\bar x_0(3x_0-2)(4x_0-1)\ln{\bar x_0}+3x_0(3x_0-1)(4x_0-3)\ln{x_0}  -14(1-5x_0\bar x_0)
%-3(x_0-1)(3x_0-2)(4x_0-1)\ln{\bar x_0}+3x_0(3x_0-1)(4x_0-3)\ln{x_0}-14(1+5x_0(x_0-1))
\Big]\nonumber\\
  &&%-\frac{5(2+9x_0(x_0-1))}{3x_0\bar x_0(1+5x_0(x_0-1))}
  -\frac{5(2-9x_0\bar x_0)}{3x_0\bar x_0(1-5x_0\bar x_0)}\frac{ \langle \textbf{q}^2\rangle_V}{M^2}\Bigg\}\,.
\end{eqnarray}
\end{subequations}

\subsection{Numerical Results}

In this subsection, we would like to show that the relativistic corrections to the LCDAs are just as important as the radiative corrections by showing the numerical results of the inverse moments and first two Gegenbauer moments.

In order to get the numerical results, we estimate $\langle {\bf q}^2\rangle_{P,V}$ by using the Gremm-Kapuskin (G-K)
 relation \cite{Gremm:1997dq}
\begin{eqnarray}
  m_{B_c}=m_b+m_c+\frac{\langle\vec q^2\rangle}{2m_b}+\frac{\langle\vec q^2\rangle}{2m_c}\,,
\end{eqnarray}
which is deduced by implementing the equations of motion from the leading order NRQCD Lagrangian.

The inverse moments and first two Gegenbauer moments of $B_c,J/\psi,\Upsilon$ are considered, the parameters are collected in Tab.~\ref{parameters}. The values of $\langle\textbf{q}^2\rangle_P$ for $B_c$, $\langle\textbf{q}^2\rangle_V$ for $J/\psi$ and $\Upsilon$ deduced from G-K relation are also listed.
Numerical results for the inverse moment and Gegenbauer moments can be calculated with these input parameters, we have
used the NDR scheme with $\Delta=0$. They are shown in Tab.~\ref{tab:moments}.
%%%%%%%%%%%%%%%
%%%%%%%%%%%%%%%

\begin{table}[!h]
\caption{Input parameters }\label{parameters}
\begin{tabular}{|c|c|c|c|c|c|}\hline
 & Meson mass  &  Quark mass & $\alpha_s$   & $\mu$&  $\langle\textbf{q}^2\rangle$  \\ \hline
 $J/\psi$ & 3.10 \rm{GeV} &$m_c=1.28$~GeV &$\alpha_s(2m_c)=0.26$ &  $ 2.56  ~\rm{GeV}$ &  $\langle\textbf{q}^2\rangle_V=0.69 ~\rm{GeV^2}$  \\
\hline
$B_c$&  6.27 \rm{GeV}    &$m_c=1.28$~GeV; $m_b=4.18$~GeV &$\alpha_s(m_b+m_c)=0.21$ & $ 5.46 ~  \rm{GeV}$  & $\langle\textbf{q}^2\rangle_P=1.59~ \rm{GeV^2}$     \\
\hline
$\Upsilon$ & 9.46  \rm{GeV} &$m_b=4.18$~GeV &$\alpha_s(2m_b)=0.19$ &  $ 8.36  ~ \rm{GeV}$  & $\langle\textbf{q}^2\rangle_V=4.6~\rm{GeV^2}$  \\
\hline
\end{tabular}
\end{table}

%%%%%%%%%%%%%%%
%%%%%%%%%%%%%%%

%%%%%%%%%%%%%%%
%%%%%%%%%%%%%%%
\begin{table}[!h]
\caption{Inverse moment and Gegenbauer moments of LCDAs for the $B_c$, $J/\psi$ and $\Upsilon$ states. Leading-order results,    one-loop QCD radiative corrections, relativistic corrections and the total results are shown, respectively.  } \label{tab:moments}
\begin{tabular}{|c|c|c|c|c|}\hline
$\langle x^{-1}\rangle $&  LO                            &   LO + $\alpha_s$                & LO+ $v^2$   & LO + $\alpha_s$+ $v^2$     \\ \hline
$B_c$& 1.31         &  1.51          &   1.53        &  1.73            \\
$J/\psi$& 2.00            &  2.31          &   2.28        &  2.59            \\
$\Upsilon$& 2.00            &  2.22          &   2.18        &  2.40         \\
 \hline\hline
$a_1 $&  LO                            &   LO + $\alpha_s$                & LO+ $v^2$   & LO + $\alpha_s$+ $v^2$     \\ \hline
$B_c$& 0.89            &  0.67          &   0.53        &  0.32            \\
 \hline\hline
$a_2 $&  LO                            &   LO + $\alpha_s$                & LO+ $v^2$   & LO + $\alpha_s$+ $v^2$     \\ \hline
$B_c$& 0.24            &  0.11          &   -0.21        &  -0.34            \\
$J/\psi$& -0.58            &  -0.26          &   -0.17        &  0.14            \\
$\Upsilon$& -0.58            &  -0.35          &   -0.33        &  -0.09            \\
\hline
\end{tabular}
\end{table}
%%%%%%%%%%%%%%%
%%%%%%%%%%%%%%%

\begin{figure}[!htb]
\centering {
\includegraphics[scale=0.6]{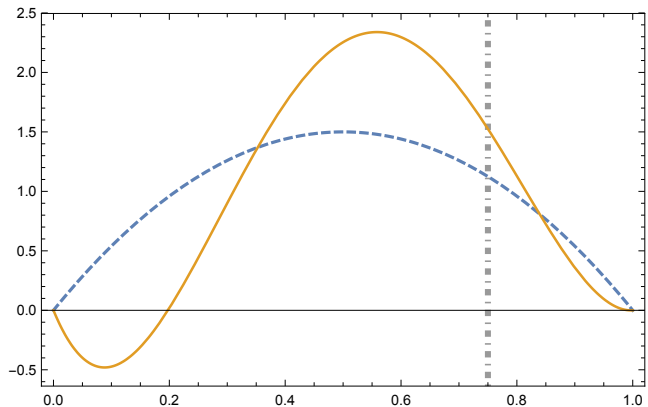}}
\caption{LCDA $\hat\phi_P(x)$ for the $B_c$ meson. The dashed line
denotes the asymptotic form $\phi(x)=6x(1-x)$, while the dot-dashed
line denotes the position $x_0=m_b/(m_b+m_c)=0.77$. The solid line is
obtained by including two Gegenbauer moments derived in
Eqs.~(\ref{mm:a1}a) and (\ref{mm:a2}a).}  \label{diagram:BcLCDA}
\end{figure}

Tab.~\ref{tab:moments} shows that the relativistic corrections to the inverse moments and first two Gegenbauer moments are comparable in magnitude with the radiative corrections, if not more important. Note that the value of $\langle {\bf q}^2\rangle_{P,V}$ deduced from the G-K relation depend on the masses of mesons and quarks. The heavy quark masses adopted here are the running masses $m_Q~(\mu=m_Q)$ $(Q=c,b)$ in $\overline{\textrm{MS}}$ scheme, which can be converted to the ``pole masses''
$m^{\textrm{pole}}_c=1.67$~GeV and $m^{\textrm{pole}}_b=4.78$~GeV, different choices of quark masses can lead to different values of $\langle {\bf q}^2\rangle_{P,V}$.
Some researches on values of $\langle\textbf{q}^2\rangle_V$ for $J/\psi$ and $\Upsilon$ based on potential model can be found in Ref.~\cite{Bodwin:2007fz} and Ref.~\cite{Chung:2010vz}. In Fig.~\ref{diagram:BcLCDA}, the LCDA $\hat\phi_P$ for the $B_c$ meson is plotted.  In this figure, the dashed line
denotes the asymptotic form $\phi(x)=6x(1-x)$, while the dot-dashed
line denotes the position $x_0=m_b/(m_b+m_c)=0.77$. The solid line is
obtained by including two Gegenbauer moments derived in
Eqs.~(\ref{mm:a1}a) and (\ref{mm:a2}a).

%%%%%%%%%%%%%%%%%%%%%%%%%%%%%%%%%%%%%%%%%%%%%%%%%%%%%%%%%%%%%%%%%%%%%%%%%%
%%%%%%%%%%%%%%%%%%%%%%%%%%%%%%%%%%%%%%%%%%%%%%%%%%%%%%%%%%%%%%%%%%%%%%%%%%

%%%%%%%%%%%%%%%%%%%%%%%%%%%%%%%
\section{Conclusion}\label{sect:Summary}
%%%%%%%%%%%%%%%%%%%%%%%%%%%%%%%

A high energy process may involve several perturbative scales. It is very important to handle these different scales, since QCD radiative corrections will induce large logarithms $\alpha_s^n\ln^{n} m_Q^2/s$.
In collinear factorization, one can relegate   the nonperturbative degrees of freedom into  LCDAs, while the logarithms  $\alpha_s^n\ln^{n} m_Q^2/s$ can be handled using the renormalization group equation, running from the scale $\sqrt{s}$ down to $m_Q$.  For the heavy quarkonium and $B_c$ system,
the refactorization scheme allows one to further reduce the nonperturbative inputs into only a  few  NRQCD matrix elements.
LCDAs of  heavy quarkonia and $B_c$ mesons are known at   next-to-leading order (NLO) in  the strong coupling constant $\alpha_s$ and  at  leading order  in  the velocity expansion.

In this paper, we have  calculated the relativistic corrections of
three twist-2 LCDAs    for the S-wave $B_c$ mesons. The
corresponding results for heavy quarkonia such as  $J/\psi$ and
$\Upsilon$ can be easily deduced by setting $m_b=m_c$. With the
results for these relativistically  corrected LCDAs, we have
studied  a few  inverse/Gegenbauer  moments. We find that the
relativistic corrections are comparable with the next-to-leading
order radiative  corrections.

%%%%%%%%%%%%%%%%%%%%%%%%%%%%%%%
\section*{Acknowledgement}
%%%%%%%%%%%%%%%%%%%%%%%%%%%%%%%

The authors are grateful to Prof.~Hsiang-nan Li and Prof.~Xin Liu for useful discussions.
This work is supported  in part   by National  Natural
Science Foundation of China under Grant
 No.11275263, 11575110, 11635009, 11655002,  Natural  Science Foundation of Shanghai under Grant  No.~15DZ2272100 and No.~15ZR1423100, and  by    Key Laboratory for Particle Physics, Astrophysics and Cosmology, Ministry of Education.

\end{document}